\documentclass[12pt, reqno]{amsart}
\usepackage{amsmath, amsthm, amscd, amsfonts, amssymb, graphicx, color}
\usepackage[bookmarksnumbered, colorlinks, plainpages]{hyperref}
\usepackage{braket}
\usepackage{algorithm}
\usepackage{algpseudocode}

\theoremstyle{definition}

\theoremstyle{remark}

\numberwithin{equation}{section}

\newcommand{\op}[2]{|#1\rangle \langle#2|}

\begin{document}
\setcounter{page}{1}

\centerline{}

\centerline{}


\title[Classical-Quantum Rainbow Table Attack on Long Passwords]{A Hybrid Classical-Quantum Rainbow Table Attack on Human Passwords}

\author[MA. Khajeian]{MA. Khajeian$^1$}

\address{$^{1}$ School of Engineering Sciences, College of Engineering, University of Tehran}
\email{\textcolor[rgb]{0.00,0.00,0.84}{khajeian@ut.ac.ir}}

\begin{abstract}
Long, human-generated passwords pose significant challenges to both classical and quantum attacks due to their irregular structure and large search space. In this work, we propose an enhanced classical–quantum hybrid attack specifically designed for this scenario. Our approach constructs rainbow tables using dictionary-based password generation augmented with transformation rules that better capture real-world user behavior. These tables are organized into buckets, enabling faster lookup and reduced space complexity.
For the search within each bucket, we employ a distributed exact variant of Grover’s algorithm. This method provides deterministic success and significantly lower circuit depth, enhancing robustness against noise—particularly depolarizing errors common in near-term quantum devices.
Overall, our hybrid framework improves the efficiency and practicality of password recovery for long, human-readable passwords in realistic adversarial settings.
\end{abstract} \maketitle

\section{Introduction}
The growing use of lengthy, user-created passwords has weakened the effectiveness of conventional password-cracking techniques. Such passwords, typically built from recognizable words or easy-to-remember sequences, often evade detection by traditional brute-force methods and standard rainbow table attacks.

In addition, the anticipated rise of quantum computing introduces new challenges and opportunities in password recovery. Although Grover’s algorithm offers a quadratic speedup for unstructured search, its direct application remains impractical in many real-world scenarios due to its high circuit depth and probabilistic nature.

To address these challenges, we revisit and extend the use of structured rainbow tables for human-generated passwords. Building on prior work that introduces a “smart dictionary” using dictionary generators and transformation rules~\cite{zhang2017improved}, we adopt a structured approach. Passwords are broken down into components governed by composition patterns. This modeling captures human tendencies and enables compact, realistic coverage of the password space. As a result, precomputation becomes feasible even for long passwords.

Based on this foundation, we propose a hybrid classical–quantum password cracking framework with two key innovations:

\begin{itemize}
    \item We partition the smart dictionary-derived rainbow table into buckets based on structural similarity or indexing heuristics. This reduces space complexity and enables fast, targeted lookups—particularly suitable for quantum memory access.
    \item We use a distributed exact Grover variant~\cite{zhou2023distributed} to perform quantum search within each bucket. This method offers deterministic success and lower quantum circuit depth, making it more robust to depolarizing noise.
\end{itemize}

This paper is organized as follows: Section~\ref{sec:background} covers fundamentals, Section~\ref{sec:methodology} presents our hybrid quantum-classical approach, Section~\ref{sec:results} shows experimental outcomes, with implementation details and conclusions following in Sections~\ref{sec:code} and~\ref{sec:conclusion}.

\section{Background}
\label{sec:background}
This section covers the rainbow table attack method, human password pattern analysis, and a modified Grover quantum search approach.

\subsection{Review of Rainbow Table Attack}
Rainbow tables are a time–memory trade-off technique used to invert cryptographic hash functions, particularly for password recovery. Originally introduced by Oechslin~\cite{oechslin2003making}, they allow attackers to precompute chains of hash outputs while storing only the first and last elements of each chain. This approach significantly reduces memory usage while enabling efficient lookups during the attack phase.

\subsubsection{Rainbow Table Structure}

A rainbow table is composed of multiple chains, where each chain represents a sequence of alternating hash and reduction operations. The construction begins with a starting plaintext $x_0$, which is hashed to produce $y_0 = H(x_0)$. This hash is then reduced using a reduction function $R_1$ to obtain the next plaintext $x_1 = R_1(y_0)$. The process continues for $t$ steps, alternating between hash and reduction functions:

\[
    x_0 \xrightarrow{H} y_0 \xrightarrow{R_1} x_1 \xrightarrow{H} y_1 \xrightarrow{R_2} \cdots \xrightarrow{R_t} x_t
\]

Only the starting plaintext $x_0$ and the final output $x_t$ are stored, forming the $(x_0, x_t)$ pair that represents the chain. To minimize chain collisions—where distinct inputs lead to the same endpoint—each reduction step uses a different function $R_i$.

\subsubsection{Table Generation and Parameters}
The effectiveness of rainbow tables depends on several key parameters. The \textit{chain length} $t$ determines how many reduction steps are applied per chain. Longer chains reduce the number of required chains but increase lookup time. The \textit{number of chains} $m$ affects the table’s coverage—more chains improve success rates but require more storage.

\textit{Reduction functions} $R_1, R_2, \dots, R_t$ convert hash outputs back into candidate plaintexts.

The generation process is illustrated in Algorithm~\ref{alg:ord_rainbow_generation}.

\begin{algorithm}[H]
\caption{Ordinary Rainbow Table Generation}
\label{alg:ord_rainbow_generation}
\begin{algorithmic}[1]
\Require Dictionary of starting plaintexts, chain length $t$
\Ensure Rainbow table entries

\ForAll{$S_{start}$ in dictionary}
    \State $S \gets S_{start}$
    \For{$i \gets 1$ to $t$}
        \State $y \gets H(S)$ \Comment{Apply hash function}
        \State $S \gets R_i(y)$ \Comment{Apply reduction function $R_i$}
    \EndFor
    \State Store pair $(S_{start}, S_{end})$ in table
\EndFor
\end{algorithmic}
\end{algorithm}

\subsubsection{Rainbow Table Lookup Process}
To look up a key in a rainbow table, the following procedure is used: First, apply the reduction function \( R_{n-1} \) to the ciphertext and check if the result matches any endpoint in the table. If a match is found, the corresponding chain can be reconstructed using the starting point. If no match is found, the process continues by applying \( R_{n-2} \) followed by \( f_{n-1} \) to check if the key appears in the second-to-last column of the table. This process is repeated iteratively, applying \( R_{n-3}, f_{n-2}, f_{n-1} \), and so on. The total number of calculations required is \( \frac{t(t-1)}{2} \) \cite{oechslin2003making}.

\subsection{Modeling Human Passwords}

Zhang et al.~\cite{zhang2017improved} proposed an improved rainbow table attack targeting long, human-readable passwords. Traditional rainbow tables use fixed dictionaries and uniform reduction functions, which fail to capture the irregular patterns found in real-world passwords. To address this, the authors introduced a \textit{smart dictionary} generated from high-frequency words and substrings in leaked datasets, combined with human-like \textit{transformation rules}.

These transformations include capitalizing letters, appending digits or years, using leetspeak substitutions (e.g., “o” to “0”), and inserting symbols. The result is a compact yet realistic password candidate set that reflects common user behavior without expanding the search space excessively.
In their evaluation, the improved method achieved a success rate of 83\% when recovering SHA1 hashes, demonstrating its effectiveness in cracking long, human-structured passwords.

\subsection{Distributed Exact Grover's Algorithm}

The Distributed Exact Grover's Algorithm (DEGA) \cite{zhou2023distributed} partitions the original $n$-qubit search problem into $\lfloor n/2 \rfloor$ subfunctions $\{g_i\}_{i=0}^{\lfloor n/2 \rfloor-1}$. Each subfunction $g_i$ is derived from the target Boolean function $f(x)$ by fixing selected groups of input bits.

For $i \in \{0,1,\ldots,\lfloor n/2 \rfloor-2\}$, the first $2i$ bits and the last $(n-2(i+1))$ bits of $x$ are fixed. This yields $2^{n-2}$ subfunctions $f_{i,j}: \{0,1\}^2 \rightarrow \{0,1\}$ defined as
\[
f_{i,j}(m_i) = f(y_{j,0} \cdots y_{j,2i-1} \, m_i \, y_{j,2i} \cdots y_{j,n-3}),
\]
where $m_i \in \{0,1\}^2$ is the variable input and $y_j$ enumerates all $(n-2)$-bit configurations. Each $g_i$ is then constructed as
\begin{equation}
\label{eq:g_i}
g_i(m_i) = \text{OR}\left(f_{i,0}(m_i), f_{i,1}(m_i), \ldots, f_{i,2^{n-2}-1}(m_i)\right),
\end{equation}
with
\[
\text{OR}(x) = 
\begin{cases}
1, & |x| \geq 1 \\
0, & |x| = 0
\end{cases}
\]
and $|x|$ denoting the Hamming weight.

For the final case $i = \lfloor n/2 \rfloor - 1$, the first $2i$ bits of $x$ are fixed, and the remaining $n-2i$ bits constitute the variable input $m_i$. The subfunction is given by
\[
f_{i,j}(m_i) = f(y_{j,0} \cdots y_{j,2i-1} \, m_i),
\]
where the dimensionality of $m_i$ depends on the parity of $n$:
\[
n-2i = 
\begin{cases}
2, & \text{if } n \text{ is even} \\
3, & \text{if } n \text{ is odd}
\end{cases}.
\]
Then $g_i$ is defined as
\begin{equation}
\label{eq:g_final}
g_i(m_i) = \text{OR}\left(f_{i,0}(m_i), f_{i,1}(m_i), \ldots, f_{i,2^{2i}-1}(m_i)\right).
\end{equation}

This partitioning guarantees that the global target state $\tau$ can be exactly reconstructed from the solutions of the subfunctions $\{g_i\}$, with each $g_i$ isolating a distinct segment of $\tau$.

Consequently, we obtain $\lfloor n/2\rfloor$ subfunctions $g_i(m_i)$ based on $f(x)$ and the parameter $n$, for $i \in \{0,1,\ldots,\lfloor n/2 \rfloor - 1\}$. We assume that each Oracle $U_{g_i}$ can be efficiently constructed. For $i \in \{0,1,\ldots,\lfloor n/2 \rfloor -2\}$, the Oracle is defined as
\begin{equation}
\label{g_i_even}
U_{g_i(x)}: \lvert x\rangle \rightarrow (-1)^{g_i(x)} \lvert x\rangle,
\end{equation}
where $x \in \{0,1\}^2$, and $\tau_i$ is the unique input satisfying $g_i(\tau_i) = 1$.

For the final subfunction $g_i: \{0,1\}^{n - 2i} \rightarrow \{0,1\}$, we distinguish two cases. If $n$ is even, $g_i$ has a 2-bit input and the same Oracle form as in Eq.~\eqref{g_i_even} applies. If $n$ is odd, a phase Oracle is used:
\begin{equation}
R_{g_i(x)}: \lvert x\rangle \rightarrow e^{i\phi \cdot g_i(x)} \lvert x\rangle,
\end{equation}
where $x \in \{0,1\}^3$, $i = \lfloor n/2 \rfloor - 1$, and $\tau_i$ satisfies $g_i(\tau_i) = 1$. The phase $\phi$ is given by
\begin{equation}
\label{eq:dega_phi}
\phi = 2\arcsin\left( \frac{\sin\left( \frac{\pi}{4J+6} \right)}{\sin\theta} \right),
\quad J = \left\lfloor \frac{\pi/2 - \theta}{2\theta} \right\rfloor,
\quad \theta = \arcsin\left( \sqrt{\frac{1}{2^3}} \right).
\end{equation}

\begin{algorithm}
\caption{Distributed Exact Grover's Algorithm}
\label{alg:dega}
\begin{algorithmic}[1]
\Require The number of qubit $n\geq2$; Oracle function $f:\{0,1\}^n \rightarrow \{0,1\}$ where $f(x)=0$ for all $x\in\{0,1\}^{n}$ except $\tau$, for which $f(\tau)=1$; $\lfloor n/2 \rfloor$ subfunctions $g_i(x)$ as in Eq.~(\ref{eq:g_i}) and Eq.~(\ref{eq:g_final}), generated according to $f(x)$ and $n$, where $i \in \{0,1,\cdots, \lfloor n/2 \rfloor - 1\}$.
\Ensure Target state $\ket{\tau}$ with certainty
\State Apply $H^{\otimes n}$ to obtain $\displaystyle |\psi\rangle = \frac{1}{\sqrt{2^n}} \sum_{x \in \{0,1\}^n} |x\rangle$
\For{$i = 0$ to $\lfloor n/2 \rfloor - 2$}
    \State Apply $G_i = -H^{\otimes 2} U_0 H^{\otimes 2} U_{g_i(x)}$
    \State where $U_0 = I^{\otimes 2} - 2(\op{0}{0})^{\otimes 2}$ and $U_{g_i(x)} = I^{\otimes 2} - 2\op{\tau_i}{\tau_i}$
\EndFor
\State Let $i={\lfloor n/2 \rfloor - 1}$
\If{$n$ is even}
    \State Apply $G_i$ to final pair of qubits
\Else
    \State Apply $L_i = -H^{\otimes 3} R_0 H^{\otimes 3} R_{g_i(x)}$ twice
    \State where $R_0 = I^{\otimes 3} + (e^{i\phi} - 1)(\op{0}{0})^{\otimes 3}$ and $R_{g_i(x)} = I^{\otimes 3} + (e^{i\phi} - 1)\op{\tau_i}{\tau_i}$ \Comment{See Eq.~\eqref{eq:dega_phi}}
\EndIf
\State Measure all qubits in the computational basis to obtain $\tau$
\end{algorithmic}
\end{algorithm}

The Distributed Exact Grover procedure is described in Algorithm~\ref{alg:dega}. To verify the correctness of the Distributed Exact Grover's Algorithm (DEGA), it suffices to show that Algorithm \ref{alg:dega} yields the target index string $\tau \in \{0, 1\}^n$ exactly \cite{zhou2023distributed}.

\section{Methodology}
\label{sec:methodology}

Our methodology builds upon the approach developed by \cite{zhang2017improved}, which demonstrated that human-chosen passwords exhibit predictable patterns regardless of length requirements. Based on this foundation, we introduce our improved attack framework.

\subsection{Dictionary Generators}
Our implementation adopts the dictionary generator framework of \cite{zhang2017improved} to construct optimized rainbow tables for password recovery. This methodology enables efficient processing of lengthy passwords while preserving human memorability patterns. To maximize occurrence probability, we employ three systematic collection approaches: (1) statistical analysis of compromised password datasets, (2) Markov model-generated strings that capture probabilistic character sequences, and (3) fundamental linguistic elements specific to target demographics.

The structural organization follows identifiable \textit{composition patterns}, where each pattern component serves a distinct function. For example, the "WNS" pattern comprises three elements: a word component (W) such as "pass", a numeric segment (N) like "1234", and special symbols (S) including "\$\$". This structured approach ensures comprehensive coverage of common password constructions.

\subsection{Transform Rules}
The framework incorporates configurable transformation rules that systematically modify dictionary entries to match prevalent password variations. Common transformations include case shifting operations (e.g., "pass" to "Pass"), special character substitutions (e.g., "E" to "3"), and complete string reversals (e.g., "well" to "llew"). The system's modular design permits seamless integration of additional transformation rules, enabling precise control over dictionary generation parameters while maintaining computational efficiency.

\subsection{Improved Rainbow Table Generation}
\label{sec:improved-rainbow-table}

To achieve high success rates, password recovery typically requires extremely large dictionaries, which presents significant storage challenges. While adopting rainbow table concepts can optimize storage, this approach necessitates redefining the core rainbow table computation functions to accommodate generator-set constraints.

A critical component is the \textit{reduction function} \( R: H \rightarrow P \), which maps hash values back to plaintext values. Practical implementations commonly decompose \( R(\cdot) \) into two sequential operations for improved efficiency:

\begin{enumerate}
    \item \texttt{HashToIndex($\cdot$)}: Transforms a hash value into an index or intermediate representation
    \item \texttt{IndexToPlain($\cdot$)}: Converts the index into a valid plaintext within the target domain
\end{enumerate}

The \texttt{HashToIndex} function implementation follows the design specified in Algorithm~\ref{alg:hash_to_index}, while \texttt{IndexToPlain} operates as detailed in Algorithm~\ref{alg:index_to_plain}. Using these components, we generate optimized rainbow tables through the process formalized in Algorithm~\ref{alg:smart_rainbow_generation}.

\begin{algorithm}
\caption{HashToIndex}
\label{alg:hash_to_index}
\begin{algorithmic}[1]
\Require Hash $H$, total plaintext candidates $N = |\mathcal{G}_1| \times |\mathcal{G}_2| \times \dots \times |\mathcal{G}_k|$
\Ensure Index $i \in \{0, 1, \dots, N-1\}$

\State $H_{\text{int}} \gets \text{BinaryToInteger}(H)$ \Comment{Convert hash to integer}
\State $i \gets H_{\text{int}} \bmod N$ \Comment{Map to plaintext space size}
\State \Return $i$
\end{algorithmic}
\end{algorithm}

\begin{algorithm}
\caption{IndexToPlain}
\label{alg:index_to_plain}
\begin{algorithmic}[1]
\Require Index $i$, generator set $\mathcal{G} = \{G_1, G_2, \dots, G_k\}$, composition pattern $P$, transform rules $\mathcal{R}$
\Ensure Target plaintext $T$

\State Initialize empty plaintext $T$
\For{each generator type $G_j \in P$} \Comment{Iterate by pattern order (e.g., "WNS")}
    \State $T_{\text{space}} \gets |G_j|$ \Comment{Size of generator subset (e.g., $|G_{\text{words}}| = 399$)}
    \State $n \gets \text{ComputeExtensionRatio}(\mathcal{R}, G_j)$ \Comment{e.g., $n=2$ if $\mathcal{R}$ includes case shifting}
    \State $T_{\text{ext}} \gets T_{\text{space}} \times n$
    \State $\text{subindex} \gets i \bmod T_{\text{ext}}$ \Comment{Local index within extended generator space}
    \State $g_{\text{base}} \gets G_j[\text{subindex} \bmod T_{\text{space}}]$ \Comment{Select base generator}
    \State $g_{\text{target}} \gets \text{ApplyTransform}(g_{\text{base}}, \mathcal{R})$ \Comment{Apply rules (e.g., "pass" $\to$ "P@ss")}
    \State $T \gets T \parallel g_{\text{target}}$ \Comment{Append to plaintext}
\EndFor
\State \Return $T$
\end{algorithmic}
\end{algorithm}

\begin{algorithm}
\caption{Rainbow Table Generation Using Smart Dictionary}
\label{alg:smart_rainbow_generation}
\begin{algorithmic}[1]
\Require Hash function $h$, generator types $\mathcal{G}$, composition pattern $P$, transform rules $\mathcal{R}$
\Ensure Rainbow table $\mathcal{T}$
\For{each chain}
    \State $S \gets \text{RandomStartIndex}()$
    \For{$t$ iterations} \Comment{$t$ = chain length}
        \State $T \gets \text{IndexToPlain}(S, \mathcal{G}, P, \mathcal{R})$ \Comment{Use Algorithm \ref{alg:index_to_plain}}
        \State $H \gets h(T)$
        \State $S \gets \text{HashToIndex}(H)$ \Comment{Use Algorithm \ref{alg:hash_to_index}}
    \EndFor
    \State Store $(S_{\text{start}}, S_{\text{end}})$ in $\mathcal{T}$
\EndFor
\State \Return $\mathcal{T}$
\end{algorithmic}
\end{algorithm}

\subsection{Bucket Creation}
\label{sec:bucket-creation}

Building on \cite{quan2024qiris}'s bucket concept, we constrain Grover's search space by hashing plaintext to $k$-bit integers and distributing them into buckets, achieving both endpoint distinction and tractable search complexity (Algorithm~\ref{alg:bucket_creation}).

\begin{algorithm}[H]
\caption{Bucket Creation}
\label{alg:bucket_creation}
\begin{algorithmic}[1]
\Require Plaintext $end$, $k$-bit hash function $\text{k\_bit\_hash}$
\Ensure Updated buckets structure

\State $end\_hashed \gets \text{k\_bit\_hash}(end)$
\State $bucket\_key \gets \lfloor end\_hashed / k \rfloor$ \Comment{Key $\in \{0, \dots, \lceil 2^k / k \rceil - 1\}$}
\If{$bucket\_key \notin buckets$}
    \State $buckets[bucket\_key] \gets \text{empty list} $
\EndIf
\State $buckets[bucket\_key].\text{append}(end\_hashed \bmod k)$ \Comment{Offset $\in \{0, \dots, k-1\}$}
\end{algorithmic}
\end{algorithm}

\subsection{Rainbow Table Search Using DEGA}
\label{sec:rainbow-table-dega}

In our approach, we reduce the hash to a plaintext using Algorithms~\ref{alg:hash_to_index} and~\ref{alg:index_to_plain}, then proceed to search for it within predefined buckets.
First, a classical linear search checks whether the bucket key of the target plaintext exists. If the bucket key is not found in the bucket list, the quantum search is skipped entirely, and the process immediately moves to the previous chain, saving time.
If the bucket key exists, we invoke the Distributed Exact Grover's Algorithm (DEGA) to search within the bucket. When DEGA successfully identifies the target, a linear search is performed to locate the corresponding hash, and the chain is reconstructed to retrieve the original plaintext. If DEGA fails to find the result, the process continues with the previous chain.
This process repeats until all chains are examined. If no match is found after all iterations, the algorithm concludes that the hash is not present in the rainbow table. The complete rainbow table search using DEGA is outlined in Algorithm~\ref{alg:rainbow_search_DEGA}.

\begin{algorithm}[H]
\caption{Rainbow Table Search with Distributed Exact Grover Search}
\label{alg:rainbow_search_DEGA}
\begin{algorithmic}[1]
\Require 
  \Statex \quad $H_{\text{target}}$: Target hash value to recover
  \Statex \quad $\mathcal{T}$: Rainbow table with:
    \Statex \qquad $\mathcal{T}_{\text{start}}$: Start points $[P_1, \dots, P_n]$
    \Statex \qquad $\mathcal{T}_{\text{end}}$: End hashes $[h_1, \dots, h_n]$ (precomputed $k$-bit hashes)
  \Statex \quad $\mathcal{B}$: Bucket structure (from Algorithm~\ref{alg:bucket_creation})
  \Statex \quad $\text{k\_bit\_hash}$: $k$-bit hash function
  \Statex \quad $\text{hash}$: Full cryptographic hash function
\Ensure 
  \Statex \quad Recovered plaintext $P$ or $\text{None}$ if not found

\State $H \gets H_{\text{target}}$
\For{$i \gets 1$ to $\text{max\_chain\_length}$} \Comment{Iterate over possible chain positions}
    \State $h \gets H$
    \For{$j \gets 0$ to $i-1$}
        \State $P \gets \text{IndexToPlain}(\text{HashToIndex}(h), \mathcal{G}, P_{\text{pattern}}, \mathcal{R})$
        \If{$j < i-1$}
            \State $h \gets \text{hash}(P)$
        \EndIf
    \EndFor
    \State $h_k \gets \text{k\_bit\_hash}(P)$
    \State $\text{bucket\_key} \gets h_k \div k$
    \If{$\text{bucket\_key} \notin \mathcal{B}$}
        \State \textbf{continue}
    \EndIf
    \State $\text{lookup} \gets h_k \bmod k$
    \State $\text{result} \gets \text{DistributedExactGroverSearch}(\mathcal{B}[\text{bucket\_key}], \text{lookup})$
    \If{$\text{result} = \text{True}$}
        \State $\text{idx} \gets \text{index of } h_k \text{ in } \mathcal{T}_{\text{end\_hashed}}$
        \If{$P \neq \mathcal{T}_{\text{end}}[\text{idx}]$}
            \State \textbf{continue}
        \EndIf
        \State $P_{\text{candidate}} \gets \mathcal{T}_{\text{start}}[\text{idx}]$
        \State $h_{\text{candidate}} \gets \text{hash}(P_{\text{candidate}})$
        \For{$m \gets 1$ to $\text{max\_chain\_length} - i$} \Comment{Rebuild chain}
            \State $P_{\text{candidate}} \gets \text{IndexToPlain}(\text{HashToIndex}(h_{\text{candidate}}), \mathcal{G}, P_{\text{pattern}}, \mathcal{R})$
            \State $h_{\text{candidate}} \gets \text{hash}(P_{\text{candidate}})$
            \If{$h_{\text{candidate}} = H_{\text{target}}$}
                \State \Return $P_{\text{candidate}}$
            \EndIf
        \EndFor
    \EndIf
\EndFor
\State \Return \text{None}
\end{algorithmic}
\end{algorithm}

\section{Results}
\label{sec:results}

This section analyzes our attack's quantum search phase, comparing noise resilience and success rates across Grover's algorithm variants.

\subsection{Noise Impact on Grover Variants}

\begin{figure}[H]
    \centering
    \includegraphics[width=0.7\textwidth]{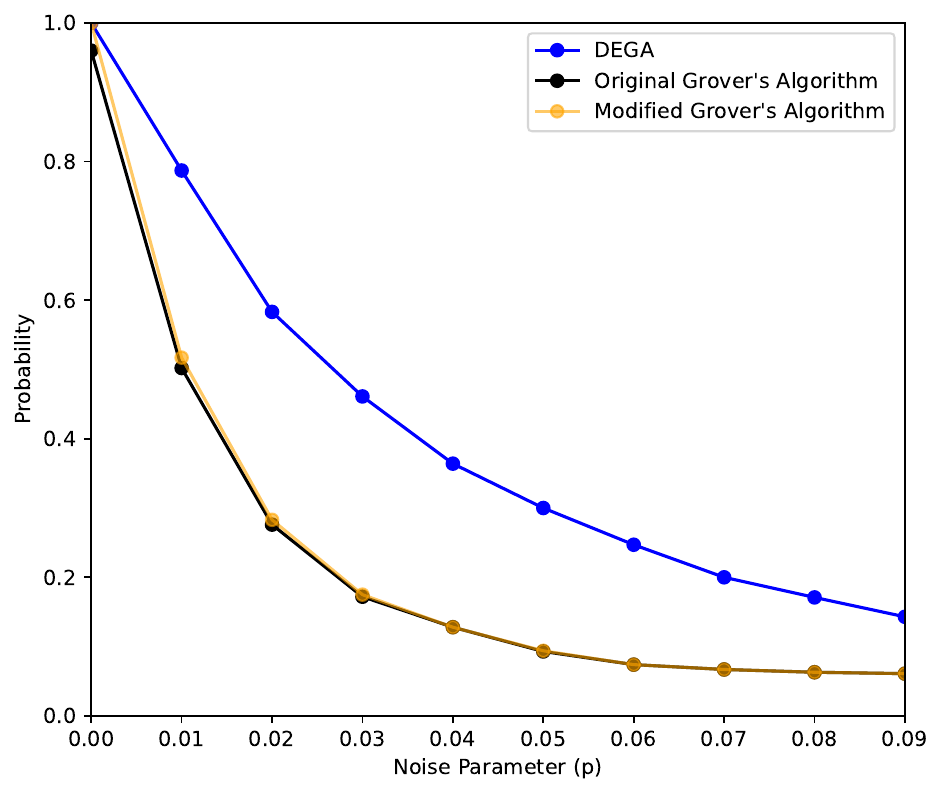}
    \caption{Probability of measuring the target state $\tau = 0011$ for the original Grover's algorithm, modified Grover's algorithm, and DEGA under depolarizing noise.}
    \label{fig:noise_comparison}
\end{figure}

To assess the robustness of Grover-based search methods under quantum noise, we evaluate their performance using a depolarizing channel model. Figure~\ref{fig:noise_comparison} compares three approaches: the original Grover's algorithm, a modified variant, and the Distributed Exact Grover's Algorithm (DEGA), testing their performance on bucket searches of size 16.

\subsection{Success Probability}

\begin{figure}[H]
    \centering
    \includegraphics[width=0.7\textwidth]{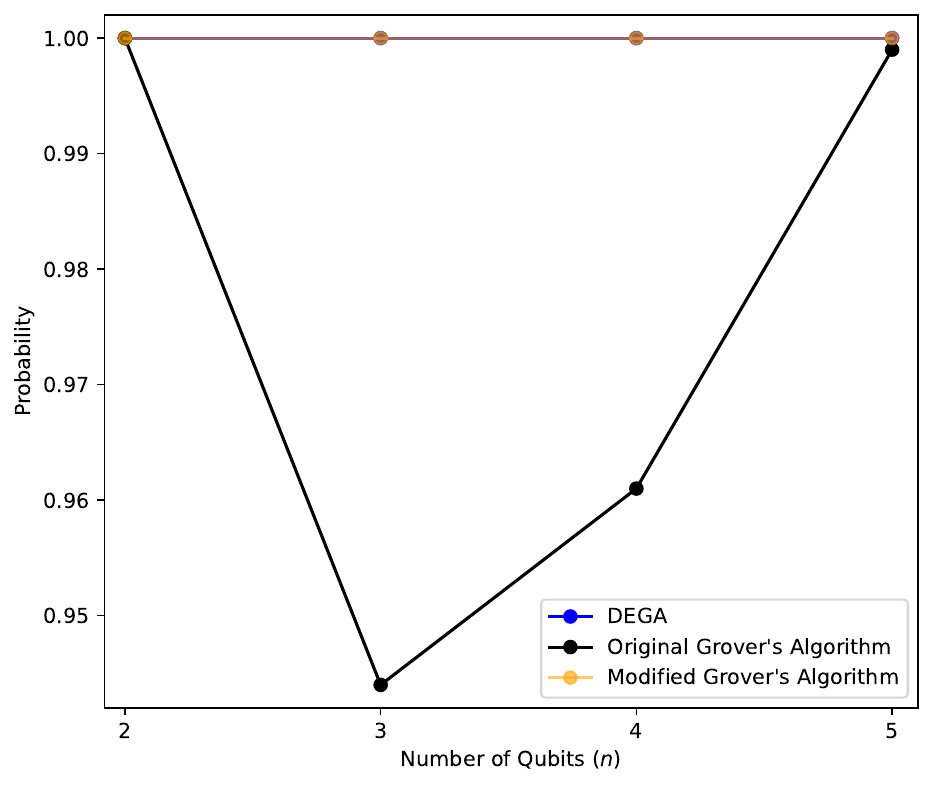}
    \caption{Probability of measuring target states $\tau \in \{11, 001, 1100, 01011\}$ across Grover variants. Both DEGA and the modified Grover's algorithm demonstrate improved exact-match performance compared to the original algorithm.}
    \label{fig:probability_comparison}
\end{figure}

Figure~\ref{fig:probability_comparison} compares the success probabilities of the original Grover's algorithm, modified Grover's algorithm, and DEGA across 2–5 qubit systems.

\section{Code Availability}
\label{sec:code}

We provide an implementation of the Distributed Exact Grover's Algorithm (DEGA) for our quantum search phase, developed using the PennyLane framework \cite{bergholm2018pennylane}. The implementation includes simulations under both ideal and noisy conditions to evaluate performance impacts on success probability. The complete source code is available at: \href{https://github.com/w0h4w4d4li/distributed-exact-grover-algorithm}{https://github.com/w0h4w4d4li/distributed-exact-grover-algorithm}

\section{Conclusion}
\label{sec:conclusion}

In this work, we introduced a classical-quantum hybrid approach to password recovery that effectively addresses the challenges posed by long, human-generated passwords. By constructing structured rainbow tables using dictionary-based generation and transformation rules, we more accurately capture real-world password patterns. These tables are efficiently organized into buckets, enabling faster and more scalable search operations.
To enhance quantum search within each bucket, we employed a distributed, exact variant of Grover’s algorithm, which provides deterministic success and reduced circuit depth. This design not only lowers overall quantum resource requirements but also improves resilience to noise in near-term quantum devices. Our results demonstrate that integrating structured rainbow tables with optimized quantum search significantly improves the efficiency and practicality of password recovery in both classical and quantum settings.

\bibliographystyle{plain}
\bibliography{main}
\end{document}